\documentclass[showpacs,preprintnumbers,amsmath,amssymb,prb,twocolumn]{revtex4}

\usepackage{graphicx}               
\usepackage{psfrag}
\usepackage{dcolumn}                
\usepackage{bm}                     

\newcommand{\ba}{\begin{array}}
\newcommand{\ea}{\end{array}}
\newcommand{\eps}{\epsilon}
\newcommand{\tr}{\textrm{ tr}}

\newcommand{\UA}{\uparrow}
\newcommand{\DA}{\downarrow}

\newcommand{\n}{\Hat{n}}

\newcommand{\phd}{^{\phantom{\dagger}}}
\newcommand{\SC}{{\cal S}}
\newcommand{\Dint}{\int {\cal D}[\SC]\;}
\newcommand{\YY}{{\cal Z}}

\begin{document}

\newcommand{\Eq}[1]{{Eq.~(\ref{#1})}}
\newcommand{\EQ}[1]{{Equation~(\ref{#1})}}
\newcommand{\av}[1]{{\left<{#1}\right>}}
\newcommand{\E}{{\textrm{e}}}
\newcommand{\note}[1]{.\newline\marginpar{\LARGE\bf!}{\bf #1}\newline}

\newcommand{\cdag}{c^\dagger}
\newcommand{\cnod}{c^{\phantom{\dagger}}}
\newcommand{\adag}{a^\dagger}
\newcommand{\anod}{a^{\phantom{\dagger}}}
\newcommand{\ctdag}{\tilde c^\dagger}
\newcommand{\ctnod}{\tilde c^{\phantom{\dagger}}}

\title{Uniform hopping approach to the FM Kondo Model at finite temperature}

\author{Winfried Koller, Alexander Pr\"{u}ll, Hans Gerd Evertz,
  and Wolfgang von der Linden}
  \affiliation{Institut f\"{u}r Theoretische Physik,
  Technische Universit\"{a}t Graz, Petersgasse 16, A-8010 Graz,
  Austria.}
\email{koller@itp.tu-graz.ac.at}

\date{November 28,  2002}

\begin{abstract}
  We study the ferromagnetic Kondo model with classical corespins
  via unbiased Monte-Carlo simulations and derive a simplified model
  for the treatment of the corespins at any temperature.
  Our simplified model captures the main aspects of the Kondo model and
  can easily be evaluated both numerically and analytically.
  It provides a better qualitative understanding of the physical features of
  the Kondo model and rationalizes the Monte-Carlo results including
  the spectral density $A_k(\omega)$ of a 1D chain with nearest neighbor
  Coulomb repulsion.
  By calculating the specific heat and the susceptibility of systems up to
  size $16^3$, we determine the Curie temperature of the 3D one-orbital
  double-exchange model, which agrees with experimental values.
\end{abstract}



\maketitle

\section{Introduction}                                  \label{sec:intro}

Manganese oxides such as
La$_{1-x}$Sr$_x$MnO$_3$  and La$_{1-x}$Ca$_x$MnO$_3$ have been attracting
considerable attention since the discovery of colossal magnetoresitance
(CMR)~\cite{proceedings98,Nagaev:book}.
These materials crystalize in the
perovskite-type lattice structure where the crystal field breaks the symmetry
of the atomic wave function of the manganese $d$-electrons. The energetically
lower  $t_{2g}$ levels are occupied by three localized electrons.
Due to a strong Hund coupling their spins are aligned, forming a
localized corespin with $S=3/2$.
The electron configuration of the Mn$^{3+}$ ions is
$t_{2g}^3e_g^1$, whereas for Mn$^{4+}$ ions the $e_g$ electron is missing. Due
to a hybridization of the $e_g$ wave function with the oxygen $2p$ orbitals,
the $e_g$ electrons are itinerant and can move from an Mn$^{3+}$ ion to a
neighboring Mn$^{4+}$ via a bridging O$^{2-}$.
The interplay of various physical ingredients
such as the strong Hund coupling of the itinerant electrons to
localized corespins, Coulomb correlations, and electron-phonon coupling leads
to a rich phase diagram including antiferromagnetic insulating, ferromagnetic
metallic and charge ordered domains.
The carriers moving in the spin and orbital background show remarkable
dynamical features~\cite{horsch99,bala02}.

Since full many-body calculations for a realistic model, including all
degrees of freedom, are not possible yet, several approximate studies of
simplified models have been performed in order to unravel individual pieces
of the rich phase diagram of the manganites.
The electronic degrees of freedom are generally treated by a Kondo lattice
model, which in the strong Hund coupling limit is commonly referred to as
the double-exchange (DE) model, a term introduced by Zener~\cite{zener51}.
In addition, the correlation of the itinerant $e_g$ electrons is well
described by a nearest neighbor (n.n.)\ Coulomb interaction.
The on-site Hubbard term merely renormalizes the already strong Hund
coupling.
For the Kondo model with quantum spins, it is still impossible to derive
rigorous numerical and analytical results.
If the S=3/2 corespins are treated classically, however, the model can be
treated by unbiased Monte Carlo techniques.
The impact of quantum spins on the electronic properties has been studied
in Ref.~\onlinecite{EdwardsI,Nolting01,Nolting03}.
It appears that quantum effects are important for S=1/2 corespins or at
$T=0$.
For finite temperature and S=3/2, classical spins present a reasonable
approximation.

Elaborate Monte Carlo (MC) simulations for the FM Kondo model with
classical $t_{2g}$ corespins have been performed by Dagotto
{\em et al.}~\cite{dagotto98:_ferrom_kondo_model_mangan,
yunoki98:_static_dynam_proper_ferrom_kondo,yunoki98:_phase},
Yi {\it et al.}~\cite{Yi_Hur_Yu:spinDE}, and by
Furukawa {\it et al.}~\cite{furukawa98,Motome_Furukawa_3dDE}.
Static and dynamical properties of the model have been determined.
These studies revealed features which have been interpreted as signatures of
phase-separation (PS).
PS has also been reported~\cite{Millis_PS} from computations based on a
dynamical mean field treatment of the DE model at $T=0$.
A phase diagram and critical exponents of the DE model have been determined
with a Hybrid MC algorithm\cite{Mayor_hybrid2001,Mayor_DE2001}.

In the manganites, the Hund coupling $J_H$ is much stronger than the
kinetic energy.
Consequently, configurations are very unlikely in which the
electronic spin is antiparallel to the local corespin.
It is therefore common practise to use $J_H=\infty$ and to
ignore antiparallel spin arrangements altogether.
This approximation yields reasonable results in the ferromagnetic regime.
Close to half-filling, however, a finite ferromagnetic Hund coupling even
enhances the antiferromagnetic ordering of the corespins.
In a previous paper~\cite{KollerPruell2002a}, we have proposed an effective
spinless fermion (ESF) model that takes effects of antiparallel $t_{2g}-e_g$
spin configurations into account via virtual excitations.
It has been demonstrated that the results of the ESF model are in excellent
agreement with those of the original Kondo model even for moderate values
of $J_H$.

In Ref.~\onlinecite{KollerPruell2002a} we also introduced the uniform
hopping approach (UHA), which replaces the influence of the random
corespins on the $e_g$ electron dynamics by an effective uniform
hopping process.
In that work, the hopping parameter was determined
by minimization of  the ground-state energy.
Essential physical features of the original model
could be described even quantitatively by UHA, while
the configuration space, and hence the numerical effort,
was reduced by several orders of magnitude.
Besides the numerical advantage, UHA also allows the derivation of
analytical results in some limiting cases.

In the present paper we extend the UHA to finite temperature.
Thermal fluctuations of the corespins are mapped to fluctuations of the
uniform hopping parameter.
In order to include entropy effects correctly, we have to determine
the density $\Gamma(u)$ of corespin states for a given hopping parameter~$u$.
The reliability of the finite-temperature UHA is scrutinized by a detailed
comparison of the results for various properties of the one-orbital DE model
with unbiased MC data.

So far, in most  MC simulations the Coulomb interaction of the $e_g$
electrons has been neglected due to its additional computational burden.
It should, however, have an important impact, particularly on phase
separation.
Moreover, at quarter filling the n.n.\ repulsion leads to the charge ordering
(CO) phase.
We have performed MC simulations for the Kondo model including a
Hubbard-like Coulomb interaction.
In these simulations, for each classical corespin configuration, the
electronic degrees of freedom are treated by Lanczos exact diagonalization.
We find that also in this case UHA yields reliable results while
reducing the computational complexity by orders of magnitude.
An excerpt of the results is given here, while a thorough discussion will
be provided elsewhere.

Also starting from an UHA-type of Hamiltonian, Millis
{\it et al.}~\cite{millis95:DE_resistivity}
claim that the bare DE model cannot even explain the right order of magnitude
of the Curie temperatures of the manganites.
This claim is, however, based on uncontrolled additional approximations.
We find that a more rigorous evaluation of UHA for a
one-orbital DE model and large 3D systems yields a Curie
temperature which is indeed close to the experimental values.
Our results for the DE model are in accord with the Hybrid MC
result~\cite{Mayor_DE2001} and with other
estimates~\cite{furukawa98,yunoki98:_phase,roeder97}.

This paper is organized as follows.
In Sec.~\ref{sec:model} the model Hamiltonian is presented and
particularities of the MC simulation are outlined.
The uniform hopping approach is discussed in Sec.~\ref{sec:UHA}.
One-dimensional applications are given in Sec.~\ref{sec:results} and compared
with MC data.
The impact of Coulomb correlations on the spectral density is discussed.
In Sec.~\ref{sec:3d} the UHA is used to calculate the phase diagram
of the DE model in 3D.
The key results of the paper are summarized in Sec.~\ref{sec:conclusion}.

\section{Model Hamiltonian and unbiased Monte Carlo}           \label{sec:model}

In this paper, we will concentrate on properties of the itinerant
$e_g$ electrons interacting with the local $t_{2g}$ corespins.
It is commonly believed that the electronic degrees of freedom are well
described by a multi-orbital Kondo lattice model
\begin{widetext}
\begin{equation}                                           \label{eq:H0}
  \hat H = -\;\sum_{i j \alpha \beta  \atop\sigma}\;
  t_{i\alpha,j\beta}\;
  \adag_{i\alpha\sigma}\,\anod_{j\beta\sigma}
  - J_H \sum_{i\alpha} \Vec{\sigma}_{i\alpha} \Vec{S}_i
 + \sum_{ij\alpha\beta} V_{i\alpha,j\beta}\;\n_{i\alpha} \n_{j\beta}
  + J'\sum_{<ij>} \mathbf S_i \cdot \mathbf S_j\;.
\end{equation}
\end{widetext}
As proposed by de Gennes~\cite{gennes60},
Dagotto {\it et al.}~\cite{dagotto98:_ferrom_kondo_model_mangan,dagotto01:review} and
Furukawa~\cite{furukawa98}, the $t_{2g}$ spins~$\mathbf S_i$ are treated
classically, which is equivalent to the limit $S\to \infty$.
The spin degrees of freedom are thus replaced by unit vectors $\mathbf S_i$,
parameterized by polar and azimuthal angles  $\theta_i$ and $\phi_i$,
respectively.
The magnitude of both corespins and $e_g$-spins is absorbed into the exchange couplings.

\EQ{eq:H0} consists of a kinetic term for the itinerant $e_g$ electrons with
transfer integrals
$t_{i\alpha,j\beta}$, where $i(j)$ are site indices, $\alpha(\beta)$
orbital indices, and $\sigma(\sigma')$ spin indices.
The transfer integrals, which are restricted to n.n.\ sites, are given
as matrices in the orbital indices $\alpha,\beta=1 (2)$ for
$x^2-y^2$ ($3z^2-r^2$) orbitals (see e.g. Ref.~\onlinecite{dagotto01:review})
\begin{equation}                                           \label{eq:hopping}
  t_{i,i+\hat z} \;=\; t \;
    \left(\ba{cc}
    0 & 0 \\ 0 & 1
    \ea \right),\;
  t_{i,i+\hat x/\hat y} =  t \;
    \left(\ba{cc}
    \tfrac{3}{4} &\mp\tfrac{\sqrt{3}}{4} \\ \mp\tfrac{\sqrt{3}}{4} & \tfrac{1}{4}
    \ea \right)\;.
\end{equation}

The overall hopping strength is $t$, which will be used as unit of energy, by setting
$t=1$. The operators
$\adag_{i\alpha\sigma} (\anod_{i\alpha\sigma})$ create (annihilate)
$e_g$ electrons at  site $x_i$ in  orbital $\alpha$ with spin $\sigma$.
The second term of the Hamiltonian describes the Hund coupling with
exchange integral $J_H$, where $\vec{\sigma}_{i\alpha}$ stands for the spin of
the electron at site $i$ in orbital $\alpha$.
The spin-resolved occupation number operator is denoted by
$\n_{i\alpha\sigma}$.
The third term describes a Coulomb repulsion, with
$\n_{i\alpha}$ being the spin-integrated density operator.
The local Hubbard interaction
is excluded from the sum, i.e. $V_{i\alpha,i\alpha}=0$,
as it effectively
merely modifies the Hund coupling $J_H$.
Finally, \Eq{eq:H0} contains a superexchange term.
The value of the exchange coupling is $J'\approx
0.02$~\cite{dagotto01:review}, accounting for the weak
antiferromagnetic coupling of the $t_{2g}$ electrons.

For strong Hund coupling $J_H\gg t$, the electronic density of states (DOS)
essentially consists of two sub-bands, a lower- and an upper 'Kondo band',
split by approximately $2J_H$.
In the lower band the itinerant $e_g$ electrons move such that their spins
are predominantly parallel to the $t_{2g}$ corespins, while the opposite is
true for the upper band~\cite{wvdl82}.
Throughout this paper, the electronic density $n$ (number of electrons per
orbital) will be restricted to $0\le n\le1$, i.e.\ only the lower Kondo band
is involved.

\subsection{Effective Spinless Fermions}

It is expedient to use the individual $t_{2g}$ spin directions $\Vec{S}_i$ as
the local quantization axes for the spin of the itinerant $e_g$ electrons at
the respective sites.
This representation is particularly useful for the $J_H\to\infty$ limit, but
also for the projection technique, which takes into account virtual processes
for finite Hund coupling.
The transformation in the electronic spin is described by a local unitary
$2\times2$ matrix $U(S_i)$ with
\begin{equation}\label{eq:LQ1}
    \vec{a}_{i\alpha} = U(S_i) \;\vec{c}_{i\alpha}\;\qquad
    \vec{c}_{i\alpha} = U^\dagger(S_i) \;\vec{a}_{i\alpha}\;,
\end{equation}
where $\vec{a}_{i\alpha}$ is a column vector
with entries $a_{i\alpha\uparrow}$ and $a_{i\alpha\downarrow}$,
respectively. The transformed annihilation operators in local
quantization are represented by $\vec{c}_{i\alpha}$.
For the creation operators we have
\begin{equation}\label{eq:LQ2}
    \vec{a}^\dagger_{i\alpha} = \vec{c}^\dagger_{i\alpha} U^\dagger(S_i)\;,
    \qquad
    \vec{c}^\dagger_{i\alpha} = \vec{a}^\dagger_{i\alpha} U(S_i)\;.
\end{equation}
The unitary matrix $U(S_i)$ depends upon $S_i$ and is chosen such that it
diagonalizes the individual contributions to the Kondo exchange
\begin{equation}\label{eq:LQ3}
  \vec{\sigma}_{i\alpha} \vec{S}_i \equiv
  \vec{a}^\dagger_{i\alpha} \;(\Vec{\Sigma}\vec{S}_i)\; \vec{a}\phd_{i\alpha}\;
  =
  \vec{c}^\dagger_{i\alpha}  \bigg(  U^\dagger(S_i)
  \;(\Vec{\Sigma}\vec{S}_i)\;      U(S_i) \bigg) \vec{c}\phd_{i\alpha}\;,
\end{equation}
with $\Vec{\Sigma}$ being the vector of Pauli matrices.
The eigenvalues of of $(\Vec{\Sigma}\vec{S}_i)$ are $\pm 1$ and the matrix of
eigenvectors is given by
\begin{equation}                                             \label{eq:U_eig}
    U(S_i) =
    \begin{pmatrix}
     c_i              & s_i  \;\E^{-i \phi_i} \\
     s_i \;\E^{i \phi_i}  & -c_i
\end{pmatrix}\;,
\end{equation}
with the abbreviations $c_j = \cos(\theta_j/2)$ and $s_j = \sin(\theta_j/2)$
and the restriction $0\le\theta_j\le\pi$.
The Kondo exchange term in \Eq{eq:LQ3} in the new representation reads
\begin{equation}                                    \label{eq:LQ4}
  \vec{\sigma}_{i\alpha} \vec{S}_i = \n_{i\alpha\UA} - \n_{i\alpha\DA}\;.
\end{equation}
The spin-integrated density operators $\hat
n_{i\alpha}$ are unaffected by the unitary transformation.
The entire Kondo Hamiltonian becomes
\begin{widetext}
\begin{equation}                                           \label{eq:H}
  \hat H = -\sum_{i j \alpha \beta  \atop\sigma \sigma'}\;
  t^{\sigma,\sigma'}_{i\alpha,j\beta}\;
  \cdag_{i\alpha\sigma}\,\cnod_{j\beta\sigma'}
  + 2 J_H \sum_{i\alpha} \n_{i\alpha\DA}
  + \sum_{ij\alpha\beta} V_{i\alpha,j\beta}\;\n_{i\alpha} \n_{j\beta}
  + J'\sum_{<ij>} \mathbf S_i \cdot \mathbf S_j\;.
\end{equation}
\end{widetext}
We have added an additional term $\hat H_c=  J_H \hat{N}$
proportional to the $e_g$-electron number $N$,
equivalent to a trivial shift of the chemical potential.

The prize to be paid for the simple structure of the Hund term is that
the modified hopping integrals $t^{\sigma,\sigma'}_{i\alpha,j\beta}$ now
depend upon the $t_{2g}$ corespins:
\begin{equation}                                      \label{eq:modihop}
  t^{\sigma,\sigma'}_{i\alpha,j\beta} \;=\;
  t_{i\alpha,j\beta}\;\big(U^\dagger(S_i)U(S_j)\big)_{\sigma,\sigma'}\;=\;
  t_{i\alpha,j\beta}\; u_{ij}^{\sigma,\sigma'}\;.
\end{equation}
The relative orientation of the $t_{2g}$ corespins at site $i$ and
$j$ enters via
\begin{equation}                                     \label{eq:u_rot}
 \begin{aligned}
    u^{\sigma,\sigma}_{i,j}(\SC) &= c_i c_j + s_i s_j \;
    \E^{i\sigma (\phi_j-\phi_i)} \hspace*{-1em} &= \cos(\vartheta_{ij}/2)\;\E^{i\psi_{ij}}\\
    u^{\sigma,-\sigma}_{i,j}(\SC) &= \sigma(c_i s_j\;\E^{-i \sigma \phi_j} -
    c_j s_i \; \E^{-i\sigma \phi_i})\hspace*{-1em} &= \sin(\vartheta_{ij}/2)\;\E^{i\chi_{ij}}
  \end{aligned}\;.
\end{equation}
These factors depend on the relative angle $\vartheta_{ij}$ of
corespins $\mathbf S_i$ and $\mathbf S_j$ and on some complex phases
$\psi_{ij}$ and $\chi_{ij}$.
It should be noticed that the modified hopping part of the Hamiltonian is
still Hermitian, because $u_{i,j}^{\sigma,\sigma'} = \big(u_{j,i}^{\sigma',
  \sigma}\big)^*$.

The advantage of the local quantization is,
as described in Ref.~\onlinecite{KollerPruell2002a}, that the energetically
unfavorable states with $e_g$ electrons antiparallel to the local $t_{2g}$
corespins can be integrated out.
This leads to the effective spinless fermion model
\begin{widetext}
\begin{equation}                                       \label{eq:Hp}
  \hat H_p = -\sum_{i,j,\alpha,\beta}
    t^{\UA\UA}_{i\alpha,j\beta}\,
    \cdag_{i\alpha}\,\cnod_{j\beta}
        - \sum_{i,j,\alpha,\beta,\alpha'}
    \frac{t^{\UA\DA}_{i\alpha',j\beta}\,t^{\DA\UA}_{j\beta,i\alpha}}
    {2J_H}\, \cdag_{i\alpha'}\cnod_{i\alpha}
    + \sum_{ij\alpha\beta} V_{i\alpha,j\beta}\;\n_{i\alpha} \n_{j\beta}
    + J'\sum_{<ij>} \mathbf S_i \cdot \mathbf S_j \;.
\end{equation}
\end{widetext}
The spinless fermion operators correspond to spin-up electrons (relative
to the {\em local} corespin-orientation) only.
The spin index has therefore been omitted.
With respect to a {\em global} spin-quantization axis, the ESF
model~(\ref{eq:Hp}) still contains contributions from both spin-up and
spin-down electrons.
The $V$-dependent contributions in the energy denominator have been ignored,
since $|V_{i\alpha,j\beta}|\ll |J_H|$.
In principle, the effective Hamiltonian also contains ``three-site'' hopping
processes.
It has been shown~\cite{KollerPruell2002a} that the three-site term has
negligible impact, and it has been ignored here.

Since each eigenvector can have an arbitrary phase,
the unitary matrix in \Eq{eq:U_eig} is not unique.
This implies that
\[
    U(S_i) =
    \begin{pmatrix}
     c_i              & s_i  \;\E^{-i \phi_i} \\
     s_i \;\E^{i \phi_i}  & -c_i
\end{pmatrix}
    \begin{pmatrix}
     \E^{i\alpha(i)}              & 0 \\
     0  & \E^{i\beta(j)}
\end{pmatrix}
\;
\]
also diagonalizes the Kondo term.
The additional phase factors modify the hopping integrals of the
spin-up channel as
\begin{equation}
  \begin{aligned}
    u^{\UA\UA}_{i,j}(\SC) &= \Big(c_i c_j + s_i s_j \;
    e^{i (\phi_j-\phi_i)}\Big)\; \E^{i(\alpha(j)-\alpha(i))} \\
    &= \cos(\vartheta_{ij}/2)\;\E^{i(\psi_{ij}+\alpha(j)-\alpha(i))}\;.
  \end{aligned}
\end{equation}
Consequently, in the one-dimensional case and with open boundaries,
we can choose the local phase factors such that the n.n.\ hopping
integrals are simply given by the real numbers $\cos(\vartheta_{ij}/2)$.

\subsection{Grand Canonical Treatment}

Our model contains classical (corespins) and quantum mechanical
($e_g$-electrons) degrees of freedom.
The appropriate way to cope with this situation in statistical
mechanics is to define the grand canonical partition function as
\begin{equation}                                           \label{eq:Y}
\begin{aligned}
  \YY &= \Dint\,\tr_c\, \E^{-\beta (\hat H(\theta,\phi)-\mu \hat{N})}\\
  \Dint &=  \prod_{i=1}^L\;\Big(\int_{0}^{\pi} d\theta_i\sin \theta_i
  \int_{0}^{2\pi} d\phi_i\Big)\;,
\end{aligned}
\end{equation}
where $\tr_c$ indicates the trace over fermionic degrees of freedom at
inverse temperature $\beta$, $\hat{N}$ is the operator for the total number of
$e_g$ electrons, $L$ is the number of lattice sites, and $\mu$ stands for the
chemical potential.
Upon integrating out the fermionic degrees of freedom, we obtain
the statistical weight of a corespin configuration $\SC$
\begin{equation}                                  \label{eq:MC_weight}
    w(\SC) = \frac{\tr_c\, \E^{-\beta (\hat H(\SC)-\mu \hat{N})}}{\YY}\;.
\end{equation}

\EQ{eq:Y} is the starting point of Monte Carlo simulations for the Kondo
model~\cite{dagotto98:_ferrom_kondo_model_mangan}
where the sum over the classical spins is performed via importance sampling.
The spin configurations $\SC$ enter the Markov chain according to the
weight factor $w(\SC)$ that is computed via exact diagonalization of the
corresponding Hamiltonian in \Eq{eq:H}.
In the 1D case we have performed MC simulations in which spins in domains of
random lengths were rotated.
We have performed MC runs with 1000 measurements.
The skip between subsequent measurements was chosen to be some hundreds of
lattice sweeps reducing autocorrelations to a negligible level.

Apart from quantities that can be derived directly from the partition
function~$\YY$, we will also be interested in dynamical observables,
notably in the one-particle retarded Green's function
$\ll \anod_{i\alpha\sigma}; \adag_{j\beta\sigma}\gg_\omega$
in global spin-quantization.
This function follows from
\begin{equation}                                         \label{eq:MC_GF}
\ll a\phd_{i\alpha \sigma}; a^\dagger_{j\beta \sigma} \gg_\omega\;
= \Dint\,w(\SC)\,
   \ll a\phd_{i\alpha\sigma}; a^\dagger_{j\beta\sigma} \gg_\omega^\SC\;.
\end{equation}
The one-particle Green's function
$\ll a\phd_{i\alpha\sigma}; a^\dagger_{j\beta\sigma} \gg_\omega^\SC$,
corresponding to a particular corespin  configuration $\SC$, is determined
from the Green's function in
local spin-quantization by employing \Eq{eq:LQ1} and \Eq{eq:LQ2}
\begin{align}                                            \label{eq:LQ5}
    \ll a\phd_{i\alpha \sigma}; a^\dagger_{j\beta \sigma} \gg^\SC_\omega
    &= \;U(S_i)_{\sigma\UA} U^*(S_j)_{\sigma\UA}
    \ll c\phd_{i\alpha}; c^\dagger_{j\beta} \gg^\SC_\omega\;.
\end{align}
To arrive at \Eq{eq:LQ5}, we used the fact that in {\em local} quantization
only the spin-up channel contributes to
$\ll c\phd_{i\alpha\sigma};c^\dagger_{j\beta\sigma'} \gg^{\SC}_\omega$,
{\it i.e.} $\sigma=\sigma'=\UA$.
The spin-down channel has structures corresponding to the upper Kondo band
in which we are not interested here.
In {\em global} quantization, both spin directions contribute.
The spin-integrated Green's function reads
\begin{align}\label{eq:LQ6}
  \sum_\sigma  \ll a\phd_{i\alpha \sigma}; a^\dagger_{j\beta \sigma}
  \gg^\SC_\omega
  &= u_{ji}^{\UA\UA}(\SC)
  \ll c\phd_{i\alpha}; c^\dagger_{j\beta} \gg^\SC_\omega\;.
\end{align}
The unbiased Monte-Carlo result for the spin-integrated one-particle
Green's function is therefore determined from
\begin{equation}\label{eq:MC_GF2}
  \sum_\sigma
  \ll a\phd_{i\alpha \sigma}; a^\dagger_{j\beta \sigma} \gg_\omega
  = \Dint w(\SC) u_{ji}^{\UA\UA}(\SC)
    \ll c\phd_{i\alpha}; c^\dagger_{j\beta} \gg^\SC_\omega\;.
\end{equation}
We note that the one-particle density of states (DOS)
is independent of the choice of the spin-quantization,
because it can be determined from diagonal Green's functions in real space.

\section{Uniform Hopping Approach}                      \label{sec:UHA}

The impact of the DE mechanism on the
electronic kinetic energy can be mimicked by an {\em average} hopping
amplitude~\cite{gennes60}.
In a previous paper~\cite{KollerPruell2002a} we introduced
what we called the ''uniform hopping approach'' (UHA).
It gave strikingly good results for ground state properties.
The idea behind UHA is to replace the terms $u^{\UA\UA}_{ij}$ in the
hopping amplitude, \Eq{eq:modihop}, which correspond to
$\cos(\vartheta_{ij}/2)$ as discussed before, by a uniform value $u$.
In Ref.~\onlinecite{KollerPruell2002a} the optimal UHA parameter $u$ was
determined by minimizing the ground state energy.
Here we will extend this approach to finite temperatures by taking entropic
effects into account.

In order to introduce the finite-temperature UHA, we proceed as follows:
For a given corespin configuration characterized by the set of
angles $\{\theta_i,\phi_i\}$, we define the average $u$-value
\[
  u(\SC) = \frac{1}{N_p} \sum_{\langle i j\rangle}
  u^{\UA\UA}_{ij}(\SC)\;.
\]
Here $N_p$ is the number of n.n.\ pairs $\langle i j\rangle$.
We now replace the individual factors $u^{\UA\UA}_{ij}$ in \Eq{eq:Y} by
$u(\SC)$.
Besides $u^{\UA\UA}_{ij}$ the Hamiltonian depends on
$|u^{\sigma,-\sigma}_{ij}|^2$ and on $\mathbf S_i\cdot\mathbf S_j$,
which correspond to $\sin^2(\vartheta_{ij}/2)$ and $\cos\vartheta_{ij}$,
respectively.
As a further approximation (see below), these terms are respectively replaced
by $1-u^2(\SC)$ and $2\,u^2(\SC)-1$.

The introduction of UHA leads to the partition function
\begin{equation}
\begin{aligned}                                             \label{eq:Y1}
  \YY &= \Dint\,\int_0^1 du \,\delta(u-u(\SC))
  \tr_c\, e^{-\beta (\hat H(u)-\mu \hat{N})}\\
  &=:  \int_0^1 du \,\Gamma_{\!N_p}(u)\,\E^{-\beta\,\Omega(u)}\;.
\end{aligned}
\end{equation}
The integrand can be interpreted as the (non-normalized) thermal probability
density for the uniform hopping parameter $u$,
\begin{equation}                                      \label{eq:density_u}
    p(u\,|\,\beta) = \Gamma_{\!N_p}(u)\,\E^{-\beta\,\Omega(u)}\;.
\end{equation}
It consists of the density of corespin states $\Gamma_{\!N_p}(u)$ and the
Boltzmann factor.
The former corresponds to a density of states  and is given by
\begin{equation}
  \Gamma_{\!N_p}(u) = \Dint\,\delta(u-u(\SC))\;.
\end{equation}
It accounts for the number of different corespin configurations
(multiplicity) that give rise to the same uniform hopping amplitude~$u$.
We note that since angles $\vartheta_{ij}/2$ enter into \Eq{eq:u_rot},
this is different from the density of states of the classical Heisenberg
model.
The grand potential $\Omega(u)$
\begin{equation}                                       \label{eq:YYY}
  -\beta\; \Omega(u) = \log \tr_c\, \E^{-\beta (\hat H(u)-\mu \hat{N})}
\end{equation}
is obtained from the fermionic trace of the homogeneous version of the
Hamiltonian of \Eq{eq:Hp}, which reads 
\begin{widetext}
\begin{equation}                                       \label{eq:Hp_UHA}
  \hat H_p(u) = -u \sum_{\substack{<i,j>\\\alpha,\beta}}
    t_{i\alpha,j\beta}\,
    \cdag_{i\alpha}\,\cnod_{j\beta}
        - (1-u^2) \sum_{\substack{<i,j>\\\alpha,\beta,\alpha'}}
    \frac{t_{i\alpha',j\beta}\,t_{j\beta,i\alpha}}
    {2J_H}\, \cdag_{i\alpha'}\cnod_{i\alpha}
    + \sum_{i,j,\alpha,\beta} V_{i\alpha,j\beta}\;\n_{i\alpha} \n_{j\beta}
    + J' N_p\, (2 u^2-1)\;.
\end{equation}
\end{widetext}

The uniform hopping approach presents an enormous simplification of the
original problem.
Firstly, the evaluation of the fermionic trace simplifies considerably;
for non-interacting electrons ($V=0$) it can even be computed analytically.
Secondly, the high dimensional configuration space of the corespins
shrinks to a unit interval.
Once the density of corespin states $\Gamma_{\!N_p}(u)$ has been determined,
the integration over the corespin states can be carried out.

\begin{figure}
  \includegraphics[width=0.46\textwidth]{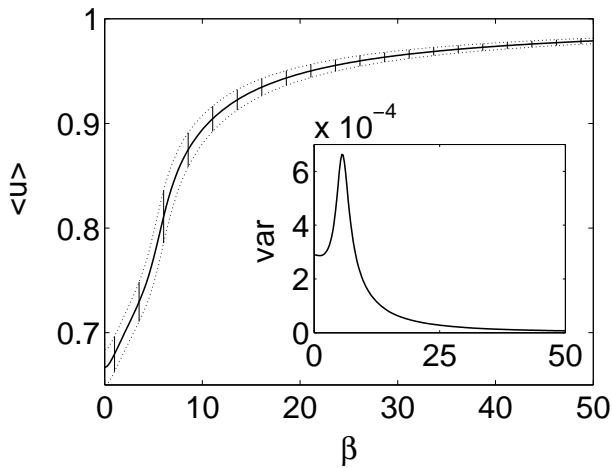}
  \caption{Mean $\left<u\right>$ (solid line)
   and standard deviation (dashed line) of  $p(u|\beta)$ versus inverse
    temperature for a $4^3$ cluster at quarter filling.
    The $\beta$-dependence  of the variance is depicted in the inset.}
  \label{av_hopping_Lx04fermi.eps}
\end{figure}

The thermal probability density $p(u\,|\,\beta)$ in \Eq{eq:density_u} contains two
competing factors.
The density of corespin states $\Gamma_{\!N_p}(u)$ peaks near $u=2/3$ and
decreases algebraically to zero as $u$ approaches the bounds of the unit
interval.
A tendency towards ferromagnetic (antiferromagnetic) order is reflected by
an exponential increase of the Boltzmann factor towards $u=1$ ($u=0$).
This factor becomes increasingly important with decreasing temperature.
In the ferromagnetic case, the combined distribution peaks,
depending on the value of $\beta$, somewhere between $2/3$ and $1$
(see Fig.~\ref{av_hopping_Lx04fermi.eps} for an illustration in 3D).
With increasing $\beta$ the peak shifts towards $u=1$.

In summary, the configuration space of the corespins is reduced to the one-parametric
space of the UHA parameter $u$.
This simplification is based on the assumption that, as far as the Boltzmann
factor is concerned, the effect of the corespins on the electrons can be
replaced by a mean effective hopping.
Fluctuations of the corespins are allowed for by the density
$\Gamma_{\!N_p}(u)$ and by fluctuations of the UHA parameter, resulting in a
finite lifetime of the quasiparticles even in the FM phase, and in a finite
bandwidth even in the AFM phase.
The density $\Gamma_{\!N_p}(u)$ takes care of the correct inclusion
of the corespin entropy, which will become crucial in the ensuing discussion.

\subsubsection*{Validity of the additional approximation}
\begin{figure}[ht]
  \includegraphics[width=0.46\textwidth]{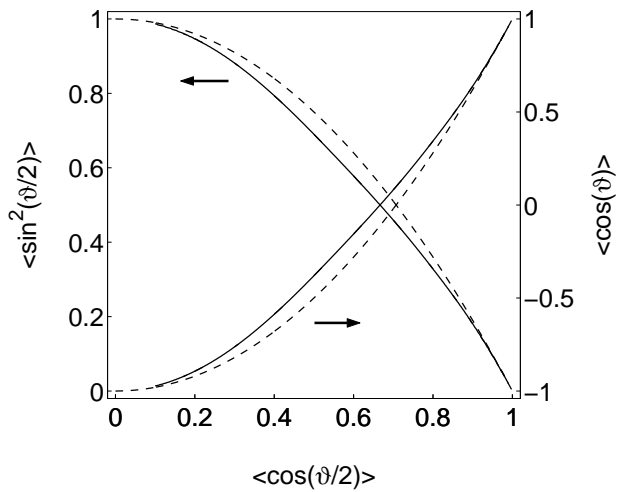}
  \caption{Average $\langle \sin^2(\vartheta/2)\rangle$ and average
    $\langle\cos \vartheta\rangle$ as a function of the average hopping
    $u=\left<\cos \vartheta/2\right>$ for a $16^3$  cluster.
    The dashed lines show the results of the 'naive' approximation explained in the text.}
  \label{sin_cos.eps}
\end{figure}

In order to assess the additional approximation introduced by the substitution of the terms
$\langle \sin^2(\vartheta/2) \rangle \approx 1-\langle\cos \vartheta/2\rangle^2 \equiv 1-u^2$ and
$\langle\cos \vartheta\rangle \approx 2\langle\cos \vartheta/2\rangle^2-1 \equiv 2u^2-1$, a Monte
Carlo simulation with random spins on a $16^3$ simple cubic (sc)
lattice has been performed. For
each spin configuration, the mean values of the functions $\cos(\vartheta/2)$, $\cos(\vartheta)$,
and $\sin^2(\vartheta/2)$ have been computed. The resulting scatter plot is depicted in
Fig.~\ref{sin_cos.eps}. Astonishingly, the data follow a unique curve and moreover they are fairly
well described by the approximation employed.

\section{UHA vs Monte Carlo in 1D}                  \label{sec:results}

In this section we scrutinize the uniform hopping approach
by a detailed comparison of its results with MC data obtained
for the original Hamiltonian \Eq{eq:Hp}.
Since the UHA affects only the treatment of the corespins, we will
restrict our attention to a one-orbital model and neglect the
degeneracy of the $e_g$ orbitals.
In this case, the Hamiltonian~\eqref{eq:Hp_UHA} simplifies to
\begin{align}                                              \label{eq:H1d}
  \hat H_p(u) =& -u\sum_{<ij>} \cdag_i\cnod_j
  - \frac{1-u^2}{2 J_H}\,\sum_i z_i\,n_i
  + V\!\!\sum_{<ij>} n_i\,n_j \nonumber \\
  &+ J'N_p\,(2u^2-1)\;,
\end{align}
where $z_i$ denotes the number of nearest neighbors of site~$i$.

For a one-dimensional chain with open boundary condition,
$\Gamma_{\!N_p}(u)$ can be calculated exactly.
For a two-site lattice we find $\Gamma_1(u)=2u\,\chi_{[0,1]}(u)$,
where $\chi_B(u)$ denotes the characteristic function of the set~$B$.
Since the relative angles of the $N_p=L-1$ nearest-neighbor pairs of a chain
of length $L$ are independent, $\Gamma_{\!N_p} (u)$ reduces to a $(N_p\!-\!1)$-fold
convolution of $\Gamma_1(u)$.
Therefore, $\Gamma_{\!N_p} (u)$ is piecewise polynomial and can be evaluated
numerically.
It can be approximated by a Gaussian, which is not
surprising because the central limit theorem applies.
In combination with the Boltzmann factor, however,
a Gaussian approximation is not good enough
because the Boltzmann factor amplifies the tails of the distribution.

\subsection{Energy distribution}

\begin{figure}[h]
  \includegraphics[width=0.95\columnwidth]{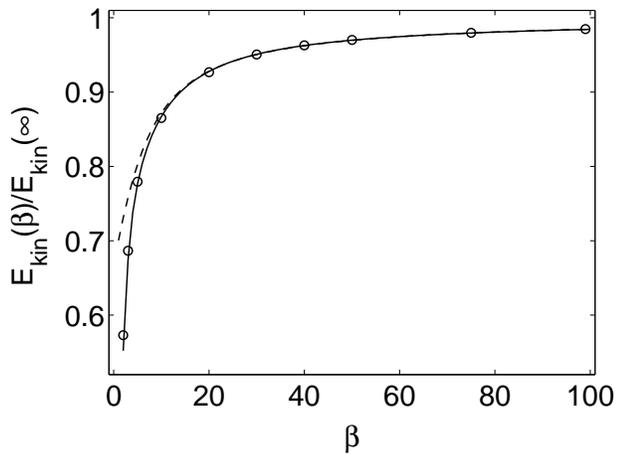}
  \caption{Kinetic energy versus $\beta$ for a
    $20$ site Kondo chain with $J_H=\infty, J'=V=0$ and $N=10$.
    The statistical errors of the Monte Carlo data (circles)
    are smaller than the marker size. The MC data are compared with
    results of UHA (solid line) and the
    canonical low-T approximations (dashed line).}
  \label{kinen_L20N10_betaX.eps}
\end{figure}

In this subsection we will compare UHA with MC results for the DE model with
$V=J'=0,\;J_H=\infty$ for a one dimensional system with one $e_g$ orbital per
site.
The Hamiltonian of \Eq{eq:H1d} reduces to a one-particle tight-binding
Hamiltonian
\begin{equation}                                        \label{eq:H1dmha}
  \hat H_p(u) = -u\sum_{<i,j>} \cdag_i\cnod_j\;,
\end{equation}
with only kinetic energy.
The hopping integral $u$ is the only remnant of the interaction with
the $t_{2g}$ corespins.
The grand potential reads
\begin{equation}
  \begin{aligned}
    -\beta \Omega(u) &= \sum_k \log(1+\E^{-\beta(\eps_k-\mu)}) \\
    &= \int dE\,\rho_L(E)\,\log(1+\E^{-\beta(E-\mu)}) \;.
  \end{aligned}
\end{equation}
where the one-particle eigenvalues $\eps_k=-2u\cos k$ depend
on $u$ and $\rho_L(E)$ denotes the tight-binding DOS of the $L$-site lattice.
$\Omega(u)$ can now be computed easily, and along with exact results for $\Gamma_{\!N_p}(u)$
we have access to the partition function and thermal quantities such as the kinetic energy.
In Fig.~\ref{kinen_L20N10_betaX.eps}, the results for the kinetic energy are
compared with those of unbiased MC simulations.
One finds an impressive agreement between the two results.
The energies are reproduced within the error bars for
all values of $\beta$.
At higher temperatures this agreement is not obvious at all, because the
corespins are strongly fluctuating.
The impact of the fluctuations seems to be well described by the UHA.

For a canonical ensemble at sufficiently low temperatures ({ canonical low-T
approximation})
one can  derive an analytical result for UHA.
To do so, the function $\Omega(u)$ is replaced by
the ground state energy of the tight-binding Hamiltonian which we write as
\begin{equation}\label{eq:low_T}
    \Omega(u) = u \;E_k\;,
\end{equation}
with a factor $E_k$ (total energy of a tight-binding system with unit hopping
amplitude) independent of $u$.
The canonical partition function then reads
\[
  Z = \Dint\, \E^{-\beta u\,E_k}\;.
\]
Since $u$ can be expressed as the average $u=\frac{1}{N_p}\,\sum
\cos(\vartheta_{ij}/2)$, the exponential function can be written as a
product of factors containing only n.n.\ spins.
In the case of a 1D chain with open boundary conditions or for a Bethe lattice,
the relative angles of neighboring spins can thus be integrated independently.
Consequently, the partition function factorizes and (up to some unimportant
constant factors) can be transformed to
\[
Z = \bigg(\int_0^1\! du \;\Gamma_1(u)\; \E^{-u \zeta} \bigg)^{N_p} =
\bigg(2\,
\frac{1 - \E^{-\zeta}(1+\zeta)}{\zeta^2} \bigg)^{N_p} ,
\]
with $\zeta = {\beta\;E_k/N_p}$.
By differentiation with respect to $\beta$, we obtain the kinetic energy
\[
E_\text{kin} =
E_k\;\frac{\zeta^2 +2\,\zeta +2 -2\,\E^{\,\zeta}}{\zeta(\zeta +1 -\E^{\,\zeta})}\;.
\]
This result is shown as a dashed line in Fig.~\ref{kinen_L20N10_betaX.eps}.
The comparison with MC results shows increasingly close agreement for
$\beta\gtrsim 10$.

\begin{figure}[h]
  \includegraphics[width=0.95\columnwidth]{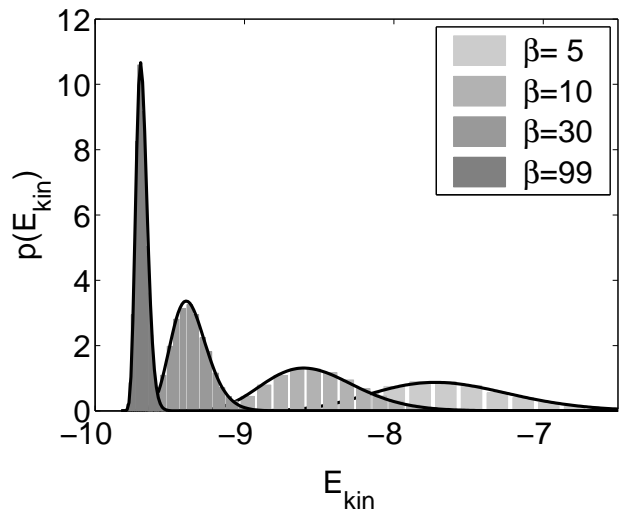}
  \caption{Probability density for the kinetic energy
    of a $16$ site DE chain at half-filling for
    various values of  $\beta$.
    The histograms are taken from unbiased Monte Carlo data.
    Solid lines represent UHA results.}
  \label{kinen_hist_L16N08_betaX.eps}
\end{figure}

The above considerations show that UHA on average correctly describes the
kinetic processes.
In order to give a more critical assessment of UHA, we study the fluctuations
of the kinetic energy.
It should be kept in mind that the motivation of UHA is to describe the mean
energy correctly.
It is thus not a priori obvious whether UHA also properly reflects its
fluctuations.
In UHA the fluctuations of the kinetic energy are
exclusively due to fluctuations of the uniform hopping parameter $u$,
that in turn is related to the relative n.n.\ angles of corespins.
In the full model, however, the relative n.n.\ angles fluctuate locally.

By sampling the contributions to the kinetic energy in a MC
simulation including local fluctuations, we obtain histograms
for the full model.
They can be compared with the statistical distribution of the kinetic
energy corresponding to the UHA density
$\Gamma_{\!N_p}(u)\, \E^{-\beta \Omega(u)}$.
The result of this comparison is depicted in
Fig.~\ref{kinen_hist_L16N08_betaX.eps}.
We find perfect agreement between MC and UHA results, revealing a
non-trivial justification of UHA.

\subsection{Spectral function and Coulomb Correlations}  \label{ssec:coulomb}

We will now comment on the influence of the n.n.\ Hubbard interaction on the
spectral density and compare MC with UHA results.
A thorough discussion of correlation effects in conjunction with the
Kondo model will be given elsewhere~\cite{KollerPruell2002c}.
We have studied a 12 site chain with open boundaries at half filling of the
effective spinless model, i.e.\ quarter filling of the original Kondo model.
In this case the implementation of the ESF model reduces the dimension of the
Lanczos basis from ${2L \choose N} = 134\,596$ to ${L \choose N} = 924$.
Additionally UHA replaces the sampling of spin configurations with a simple
scan in the UHA parameter $u$ (several $100\;000$ spin configurations in MC
versus approx.\ $20$ $u$-values in the relevant $u$-range $[0.8\,,\,1.0]$ in
UHA).
\begin{figure}
  \includegraphics[width=0.40\textwidth]{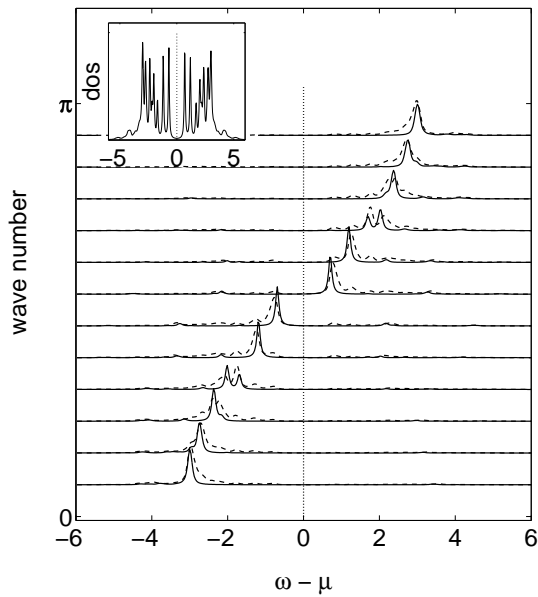}
  \caption{Spectral density of a 12-site Kondo chain at quarter
    filling ($N=6$) with $V=2$, $J'=0.02$,
    $J_H=6$, $\beta = 50$.
    Data of the Monte Carlo -- Lanczos hybrid algorithm (dashed lines)
    are compared with UHA results (solid line).
    The inset displays the DOS obtained from MC simulations.}
  \label{Ak_Lanc_L12N06V2.0.eps}
\end{figure}

Without Hubbard interaction, the system is ferromagnetic due to the DE
mechanism.
The spectral density, calculated by MC and depicted in Fig. 12 of
Ref.~\onlinecite{KollerPruell2002a}, is that of a tight-binding
model~\cite{yunoki98:_static_dynam_proper_ferrom_kondo}.
The peaks are slightly broadened due to spin fluctuations.
In UHA, through the variation of the uniform hopping amplitude~$u$, we obtain
a superposition of tight-binding bands that combine to a broadened
tight-binding band.
For the parameters $L=20$, $J_H=6$ and $J'=0.02$ and at $\beta=50$, the
average uniform hopping amplitude $\left<u\right>$ is found to be
$\left<u\right>\simeq 0.953$.
This yields a band width~$W$ of $W\simeq 3.8$ which agrees with what
we have found in MC simulations.

We now include the n.n.\ Hubbard term with $V=2$ in the ESF
model \Eq{eq:Hp}, or alternatively in the UHA Hamiltonian in \Eq{eq:Hp_UHA}.
The Monte Carlo data are obtained by resorting to a Lanczos exact
diagonalization scheme for each corespin configuration.
The fermionic trace is then evaluated by summing over enough lowest
eigenstates, such that convergence is ensured.
Details will be given elsewhere~\cite{KollerPruell2002c}.

In UHA, a $t-V$ model has to be diagonalized.
The Lanczos diagonalization for this model is  not really faster than the
diagonalization of the original model, but the configuration space is
drastically reduced, as only the parameter $u$ has to be sampled within the
unit interval $u\in [0,1]$.

Figure~\ref{Ak_Lanc_L12N06V2.0.eps} shows the spectral density derived by
both approaches.
The electronic correlation has important impact on the spectral density.
A gap appears in the middle of the original Brillouin zone at $k=\pi/2$,
indicating the doubling of the unit cell due to charge ordering.
In addition, the spectra exhibit more structure than just a simple quasi
particle peak.

This result is neither new nor surprising. The point we want to make
here is that UHA works well also for correlated electrons,
indicating that it can reliably be employed to study more sophisticated and
more realistic models for the manganites, e.g. by including correlation
effects, phononic degrees of freedom, and more orbital degrees of freedom.

\section{FM phase transition in 3D}                         \label{sec:3d}

We now apply UHA to a sc crystal and determine the Curie temperature for the bare
one-orbital DE model.
The crucial difference between the 1D and the 3D geometry
is that in the latter the relative angles of n.n.\ corespin-pairs are
in general correlated.
Therefore the correct density $\Gamma_{\!N_p}(u)$ is no longer a convolution
of the density $\Gamma_1(u)$ of a single spin-pair.

\subsection{Determination of $\Gamma_{\!N_p}(u)$}

\begin{figure}[h]
  \includegraphics[width=0.46\textwidth]{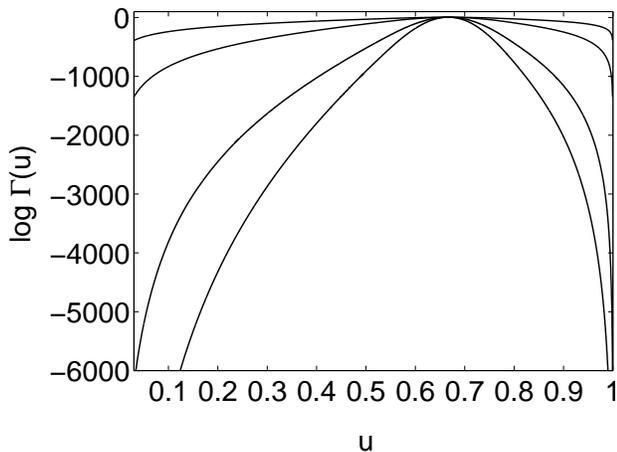}
  \caption{Density of corespin states $\Gamma_{\!N_p}(u)$ versus UHA
    parameter $u$ for a sc lattice with linear size $L_x=4$ (top),
    $L_x=6, 10$ and $L_x=12$ (bottom).
    All curves peak near $u=2/3$, as in the 1D case.}
  \label{spindos_3d_Lx.eps}
\end{figure}

In order to determine $\Gamma_{\!N_p}(u)$ for a 3D geometry, we
have to resort to numerical approaches.
We have employed the Wang-Landau algorithm~\cite{wang01:_deter}
with single spin flip updates,
which was invented for the determination of
the density of states of classical models.
Figure~\ref{spindos_3d_Lx.eps} shows the resulting density
$\Gamma_{\!N_p}(u)$ as a function of $u$ for a sc lattice with linear
dimensions $L_x=4,6,10$ and $12$.
As in the one-dimensional case, $\ln(\Gamma_{\!N_p}(u))$ diverges as
$u\to 0$ and $u\to 1$.
In fact, one can show that
\begin{equation}                     \label{eq:u_asymp}
  \ln(\Gamma_{\!N_p}(u)) \underset{u\to 1}{\longrightarrow}(L-1)\;\ln(1-u)
\end{equation}
in any spatial dimension.
This divergence has important impact on the low-temperature thermodynamic
behavior. The entropy diverges logarithmically and the specific heat has
a finite value for $T\to 0$.
The scale in Fig.~\ref{spindos_3d_Lx.eps} might appear exaggerated,
but it is actually the tiny tail close to $u=1$ which will
become important for low temperatures.

The computational effort of finite-temperature UHA is now essentially reduced
to the Wang-Landau determination of $\Gamma_{N_p}(u)$, while the integration
over $u$ to calculate various physical results takes only a small amount
of computer time.
Therefore, results can be obtained for much larger lattices than with the
conventional MC approach and, indeed, for a whole range of temperatures at
once.

\subsection{FM to PM transition at $J_H=\infty,J'=0$}

We now study the 3D DE model in the UHA.
Based on the tests of the previous section, we expect the UHA results to be
reliable also in this case.
We restrict the present discussion to the case $J_H=\infty,J'=0$.
For these parameters, only the FM and PM phases
exist~\cite{KollerPruell2002a,brink99:_DE_two_orbital}.

The trend from PM to FM can already be seen in
Fig.~\ref{av_hopping_Lx04fermi.eps},
where we show the expectation value $\left<u\right>$ of the
uniform hopping parameter and its standard deviation as a function of the
inverse temperature $\beta$ at $\mu=0$.
Already for a relatively small system, $p(u\,|\,\beta)$ is sharply
peaked.
Starting from $u=2/3$ at high temperatures, the expectation value
$\left<u\right>$ tends towards unity, {\it i.e.} FM corespins,
as $\beta \to \infty$.
From \Eq{eq:u_asymp} we find the asymptotic formula
\begin{equation}                      \label{eq:u_star}
  u^* = 1+\frac{1}{\beta \eps_k}
\end{equation}
for the position $u^*$ of the maximum of $p(u\,|\,\beta)$, where
$\eps_k$ denotes the kinetic energy per lattice site of the tight-binding
model with unit hopping parameter.
It turns out that for $\beta\gtrsim 10$, the curves for $u^*$ and
$\left<u\right>$ coincide.
Well above this temperature, near $\beta \approx 5.5$, the variance
of $p(u\,|\,\beta)$ shows a peak (see inset of
Fig.~\ref{av_hopping_Lx04fermi.eps}), indicating important fluctuations near
this temperature.
\begin{figure}[ht]
  \includegraphics[width=0.46\textwidth]{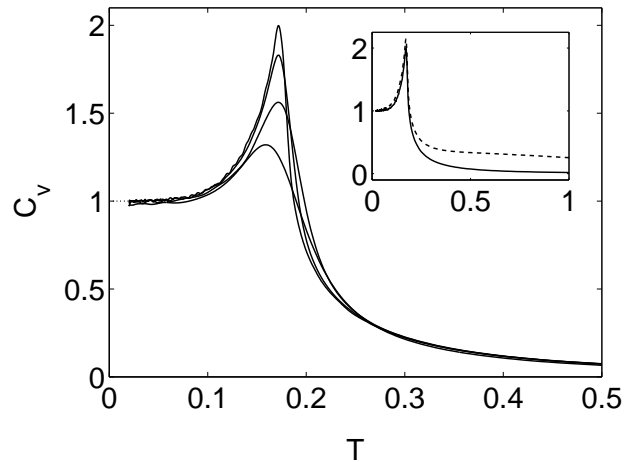}
  \caption{Specific heat per site of the sc DE model at $n=0.5$ ($\mu=0$)
    versus temperature for $L=4^3$ (bottom), $L=6^3, 10^3$ and  $L=16^3$
    (top). Parameters are $J_H=\infty,J'=0$. The results are obtained by the
    ``canonical low-T approximation'' (see text).
    In the inset, the approximate result for a $16^3$ lattice is compared
    with that of an exact grand canonical calculation (dashed line).}
  \label{Cv_3d_Lx.eps}
\end{figure}
For the determination of the Curie temperature of the DE model, we study the
specific heat $C_v$ as a function of temperature for various system sizes.
The peaks of the specific heat at quarter filling ($n=0.5$) are plotted in
Fig.~\ref{Cv_3d_Lx.eps}.
They show signs of divergence as the lattice size increases.
This indicates the presence of a second order phase transition from FM to PM.
We identify the position $T^*\simeq 0.17$ of the peak as the phase
transition temperature $T_C$ at $n=0.5$.
This value is somewhat higher than that determinded with the Hybrid MC
algorithm\cite{Mayor_hybrid2001} ($T_C\simeq 0.14$) for a $16^3$ lattice but
is better than the variational estimate\cite{Mayor_variational2001}
$T_C\simeq 0.19$.

In order to facilitate the calculation, particularly for electron fillings
different from $n=0.5$ ($\mu=0$), we consider a canonical ensemble and
replace the Boltzmann factor by $\E^{-\beta F}$. If the temperature is small
on the electronic energy scale, we can replace the electronic free energy $F$
by the ground state energy $F\simeq u E_k$. As introduced above, $E_k$
denotes the kinetic energy at $T=0$  of the tight-binding model with unit
hopping amplitude (now in 3D) for a given electron filling. This
approximation is justified because $T_C \lesssim 0.17$ is indeed small
enough. The partition function now reads 
\begin{equation}\label{eq:Z}
  Z = \int_0^1 du\,\Gamma_{N_p}(u)\,\E^{-\beta \,E_k\,u}\;.
\end{equation}
The impact of this ``canonical low-T approximation'' is illustrated in the inset of
Fig.~\ref{Cv_3d_Lx.eps}. We find that the position of the peak is not affected
at all. The only difference
to the full grand canonical result
is the longer tail at higher temperatures of the full result, which is due to
additional fluctuations of the electrons.

The specific heat approaches a constant value $C_v = 1$ as $T\to 0$.
This can be inferred from \Eq{eq:u_star}, since, for low temperatures,
the internal energy per lattice site is given by
$\eps_k\,u^*$ whose derivative with respect to temperature exactly
yields unity.
This explains the plateau of $C_v$ for $T\lesssim 0.1$.

\begin{figure}[h]
  \includegraphics[width=0.46\textwidth]{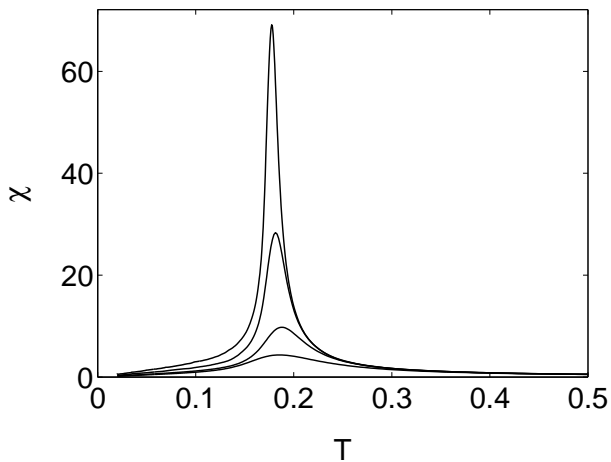}
  \caption{Magnetic susceptibility of the sc DE model
    at quarter filling ($n=0.5$) versus temperature
    for lattice sizes $4^3$ (bottom), $6^3, 10^3$ and
    $16^3$ (top). Parameters are $J_H=\infty,J'=0$.
    The results are obtained by the ``canonical low-T approximation''.}
  \label{chi_vs_T_Lx.eps}
\end{figure}

Signatures of the FM to PM phase transition should show up especially in the
magnetic susceptibility\cite{Yi_Hur_Yu:spinDE} $\chi$.
For its calculation, the density $\Gamma_{N_p}(u)$ is not sufficient because
a value $u$ of the average hopping does not determine the magnetization~$m$.
Given the conditional probability $p(m\,|\,u)$ the moments of the
magnetization are
\[
 \left<|m|^n\right> \equiv
 \frac{1}{\YY} \int_0^1 du\, \Gamma_{N_p}(u)\,
 \E^{-\beta \Omega(u)}\,M^{(n)}(u)
\]
with
\[
 M^{(n)}(u) =  \int_0^1 dm\, |m|^n p(m\,|\,u)\;.
\]
Estimates of the conditional moments $M^{(n)}(u)$ have been obtained in a
second run of the Wang-Landau algorithm.
A random walk in the space of all corespin configurations is performed whose
acceptance is controlled by $1/\Gamma_{N_p}(u)$.
An estimator of the susceptibility~\cite{binder01,binder88} is then given by
\[
  \chi = \beta\,L\,\big(\left<m^2\right>-\left<|m|\right>^2\big)\;.
\]

Figure \ref{chi_vs_T_Lx.eps} shows the susceptibility as a function
of the temperature for various lattice sizes. We observe clear signs
of a divergence near $T\simeq 0.18$ which corroborates the transition
temperature obtained from the specific heat.

The filling dependence of $T_C$ is easily determined from \Eq{eq:Z}.
Since the filling dependence only enters via $E_k$, which shows up in
combination with $\beta$, we have the simple relation
\[
  \beta_c E_k = \text{const}
\]
for the transition temperature.
Thus the Curie temperature $T_C$ is proportional to the kinetic energy
$E_k$ of the tight-binding model which, in turn, is a function of the
electron filling.
The proportionality of $T_C$ to the bandwidth has already been found based on
different approximations~\cite{Izyumov71,Nolting02,furukawa98}.
\begin{figure}[h]
  \includegraphics[width=0.46\textwidth]{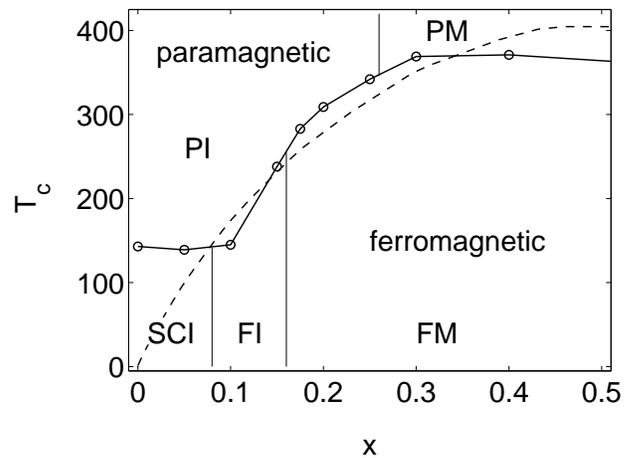}
  \caption{Curie temperature (dashed line) of the one-orbital DE
    model for a $16^3$ cluster and $t=0.2\,$eV.
    Circles and phases PM, PI, FM, FI, and SCI are experimental
    results\cite{urushibara95:_insul_la_sr_mno} for 
    La$_{1-x}$Sr$_{x}$MnO$_3$.}
  \label{LaSrMnO_phases.eps}
\end{figure}
In order to compare our calculations with experimental results, we fix
the single free parameter in the DE model, i.e., the hopping amplitude.
We choose $t=0.2$ eV, a value reasonable for the
material~\cite{dagotto01:review}.
The dashed line of Fig.~\ref{LaSrMnO_phases.eps} shows the Curie temperature
obtained from the DE model in UHA.
We find astonishingly good agreement to the experimentally observed phase
diagram of La$_{1-x}$Sr$_x$MnO$_3$ in the ferromagnetic regime.
Our result is in sharp contrast to the claim made by Millis
{\it et. al.}~\cite{millis95:DE_resistivity} that the DE
model cannot even explain the right order of magnitude of $T_C$ for the
manganites.
The reasoning of Ref.~\onlinecite{millis95:DE_resistivity} starts from
similar ideas as the UHA but is based on additional uncontrolled
approximations.
Our results for the DE model are in accord with other
estimates~\cite{furukawa98,yunoki98:_phase,roeder97}.

The experimentally observed phase diagram
shows additional phases for small concentrations:
ferromagnetic insulating (FI), paramagnetic insulating (PI) and a
spin-canting insulating (SCI) state.
These states are not accounted for in our present approach.
For a correct description, a finite value of $J'$ is important, as well
as generalizations of UHA, which will be discussed
elsewhere~\cite{KollerPruell2002c}.

\section{Conclusions}                           \label{sec:conclusion}

In this paper we have presented the uniform hopping approach (UHA) for the FM
Kondo model at finite temperature.
We have used our method to calculate the ferromagnetic to paramagnetic phase
transition temperature of the one-orbital DE model for large 3D systems.
We find that the DE model yields a Curie temperature that is comparable to
the experimental data.

The finite temperature UHA in the frame of the ESF model reduces the
numerical effort of a simulation by several orders of magnitude, while
retaining all crucial physical features.
In the example given in Sec.~\ref{ssec:coulomb}, the reduction factor is at
least $10^6$.
The key idea is to map the physics of the high dimensional configuration
space of the $t_{2g}$ corespins onto an effective one-parametric model.
The density of states entering our approach can be determined by the
Wang-Landau algorithm.
A full thermodynamic evaluation of the UHA model takes into account entropy
and fluctuations of the corespins.
Tests for 1D systems reveal that UHA results are in close agreement with
unbiased MC data for static and dynamic observables.

This reduction in numerical effort will allow us to include phononic and/or
orbital degrees of freedom in future numerical simulations in order to study
more realistic models for the manganites.

\acknowledgements

This work has been supported by the Austrian Science Fund (FWF), project
no.\ P15834-PHY.
We are indebted to W. Nolting for stimulating discussions and to
V. Mart\'{\i}n-Mayor for drawing our attention to
Refs.~\onlinecite{Mayor_DE2001,Mayor_hybrid2001,Mayor_variational2001}.


\end{document}